# THE COSMIC QUANTA PARADIGM FULFILS
# THE RELATIVISTIC MECHANICS, IMPROVES THE GRAVITATION THEORY AND ORIGINATES THE NUCLEAR FORCES.

Maurizio Michelini
ENEA, Casaccia Research Centre
Via Anguillarese 301, 00060 S.Maria di Galeria-Rome, Italy
maurizio.michelini@casaccia.enea.it

**Abstract –** This work originates from the rational conviction that physics stagnates when some old paradigm hinders the research. This is the case of the gravitational mass. Several physical problems may be explained through the new paradigm of the small cosmic quanta (whose wavelength equals the Planck's length) describing the high energy content of the void. Firstly the paradigm defines a cross-section of the elementary particles which is proportional to the respective mass. In accord with the equivalence principle this paradigm generates both the gravitation and the relativistic inertial forces. The mass-velocity dependence and the mass-energy equivalence of special relativity can be derived from it. The gravitational force between two masses is due to the differential momentum that the cosmic quanta discharge (Compton effect) on the mutually shielding masses. The present quantic gravitational theory predicts also very high figures of $G$ related to the density of the neutron stars. This result agrees with the recent reliable measurements of some huge galactic obscure bodies, whose gravity has been currently interpreted as due to some millions of Sun masses. Finally, the effect of the cosmic radiation pressure upon close nucleons is analysed, showing that the nuclear force, i.e. the strongest one, comes from the same origin of the weakest force. This determination permits us to evaluate energy, wavelength and flux of the cosmic quanta. By consequence the high energy density of the void is in accord with figures which need to explain the very high inertial forces arising from the "void" space.

## 1 – INTRODUCTION

From the beginning of the past century the astronomers made spectroscopic measurements of the light coming from remote celestial sources, whose emission lines appeared shifted toward the red. These observations showed a statistical relationship between the redshift and the apparent luminosity, which in turn may be (when possible) related to the distance of the sources.

In 1929 E. Hubble published these data showing a correlation of approximate proportionality between the redshift and the distance $s$ of the source

$$(1.1) \qquad z = (\lambda - \lambda_o) / \lambda_o \approx H s / c$$

where the constant $H$ was put between 50 and 95 km / sec.Mpc.
The recent accurate measurements by means of the Hubble spatial telescope[1,2,3] showed that $H$ is comprised between 70 and 80 km /sec.Mpc. The inverse of $H$ is linked to the age $t_o = 2/3H$ (about 10 billion years) attributed to the universe by the expanding cosmological models. In other models the universe shows a greater age, sometimes indefinite.

The astronomers know that globular star clusters show an age which, before the advent of the expanding model, was estimated higher than 10 billion years. The duration of the brown dwarfs and neutron stars, which cool very slowly, is of the order of hundred billion years. The general impression is that the universe shows a temporal extension greater than the expectation of the expanding models [4].

The interpretation Hubble gave of the redshift was quite pragmatic: light coming from the remote galaxies shows a spectral redshift which is about proportional to the source distance, without any allusion to the removal of the galaxies.

The Hubble's data were welcomed by the cosmologists, a new kind of scientists appeared after the Einstein's publication (1916) of the general theory of relativity, which proposed a new vision of the inertial-gravitational phenomena. Substantially, to explain the cosmological problem a metric structure of the universe was assumed in presence of a finite density of gravitational matter . Successively Einstein published a work in which introduced the so-called «cosmological constant», that is a repulsion force increasing with the distance, which (on the cosmic scale) balances the gravitational force, so allowing the universe to be static.

Although Einstein admitted the arbitrariness of his act , the cosmologists involved with the observations felt the opportunity of maintaining a cosmological constant different from zero[4] because the expanding models may find difficulties in reconciling the predicted age of the universe with the age of the ancient globular star clusters.

In 1922 the russian astronomer Friedmann proposed that universe were expanding. In 1927 the cosmologist George Lemaitre elaborated an expanding model which is the basis of the Big bang model. In this revolutionary atmosphere, the Hubble's data (1929) were welcomed as the proof that the universe was really expanding.

Since the wavelength $\lambda_o$ of the light emitted some billion years ago is unknown, it was assumed that the emission sequences of the atoms did not change during the cosmological times. This assumption is acceptable when the *initial* conditions of the universe result not much different from the present ones (as it happens in the static universe), but it creates troubles when referred to the expanding models which predict initial temperatures and densities *so exceptional* that the structure of matter could not emit the visible light spectrum. Anyway, at the end of the 1920s the redshift was attributed to the removal of galaxies with velocities $v \cong H s$ proportional to their distance from the observer.

After the second world war, two models described in different ways the expansion of the universe. One lost its credibility after the discovery in 1965 of the cosmic microwave background radiation, while the other was reinforced. Successively this model (Big bang) experienced serious difficulties due to the results of the observations at the periphery of the visible universe, which required a new theoretical support (inflation theory) at the end of 1970s.

It is necessary, firstly, to analyse the physical basis of the Doppler effect. In the case of the electromagnetic waves, this phenomenon shows that the spectral lines of light emitted by a source which removes with velocity $v$ respect to the observer, are displaced toward the red according to the relativistic expression

(1.2)  $$z' = (\lambda - \lambda_o) / \lambda_o = (1 + v^2/c^2)^{1/2} / (1 - v^2/c^2)^{1/2} - 1$$

which gives figures well beyond unity when the source velocity approaches $c$.

When in the 1960s some redshift *greater than unity* was observed (indicating past velocities of galaxies comparable to that of light), the relativistic expression was adopted instead of the classical one which would have required a galaxy speed greater than $c$.

Recently C.Lineweaver and T.Davis[5] pointed out that Special relativity applies to the velocity of matter respect to the inertial frames, i.e. respect to the fixed masses of the universe.

Conversely General relativity implies the *space* expansion, so very distant galaxies (which can be seen up to a distance more than 3 times[5] the Hubble radius $R = c/H$ ) may remove each other with a speed greater than $c$. In this case the redshift is nearly proportional to the source distance if we assume that $H$ is nearly constant along the distance.

In this frame, we shall try to give an explanation of the cosmological redshift no longer linked to an expanding universe, but to a *static-evolving* universe based on the new gravitational interaction acting at «local» scale. This interaction does not create forces between the clusters of galaxies, which may consequently remain at rest.

In the past the observed high redshifts gave rise to a puzzling astronomical problem. Since the absolute luminosity of the quasi-stellar objects is unknown, they may be placed within our Milky Way (gravitational redshift of dense stars) or at the periphery of the universe(cosmological redshift of protogalaxies). A solution to the problem of measuring large astronomical distances has been attempted only recently by observing the distant supernovae IA, whose distance can be evaluated as for standard candles, because their absolute luminosity is almost the same everywhere.

The interpretation of the Hubble's redshift as a Doppler effect involves that light has a constant velocity of propagation in the void space. However, strictly speaking, nobody knows if this assumption is verified *during the cosmological eras*.

The light propagation was described by the Maxwell's field equations as a function of the electric polarizability $\varepsilon$ and the magnetic permeability $\mu$ of the medium in which the field spreads

$$(1.3) \qquad\qquad c^2 = 1 / \varepsilon \mu \,.$$

According to Maxwell, the electromagnetic field propagates in the *ether*, a medium considered to be homogeneous and isotropic. Successively, the experiments by Michelson and Morley established that the existence of the ether was unsustainable and the special relativity did not retain necessary the existence of a physical medium supporting the electromagnetic field.

In other words the ether was removed, but the field description through the Maxwell's equations was entirely maintained. From this conceptual rift, a series of conjectures about the time dilation and the contraction of bodies sprang out, forgetting that purely mathematical entities, such as time and space, do not have independent physical existence. This concept was clearly expressed by H. Dingle[6] referring to the historical development of the physical science: «Galileo and Newton took observations as their starting point and used mathematics only as a tool to extract the maximum amount of information from their experiments and as a mean for expressing their new-found knowledge». After publication in 1905, a weak attention was attributed by the physicists to the Einstein's work on the electrodynamics. Viceversa it was welcomed by the mathematicians of Gottinga (Minkowski, Hilbert) which extended special relativity giving rise to the formal thinking in physical problems[7]. But the *physicist* Einstein was not convinced. In a letter to Ehrenfest (1912) Einstein wrote «You are one of a few theorists which was not despoiled of his native intelligence by the epidemic of mathematics». The scientific involution induced by the mathematical formalism leaded the research far from a physical discovery able to explain, for instance, the experimental increase of the particle mass when the velocity approaches $c$.

In general the proliferation of the mathematical formalism corresponds to the effort of turning round a physical paradigm not suspected to be unable to explain phenomena.

An old problem has to day reappeared: if the ether does not exist, to which thing the constants $\varepsilon$ and $\mu$ do refer ? The answer currently given is that these quantities refer to the *physical void*, which contains an energy density up to now not revealed.

## 2 – A NEW PARADIGM FOR THE INERTIA-GRAVITATION: THE FLUX OF COSMIC QUANTA

In the general relativity theory the space bending is due to the gravitational masses, which in the newtonian theory draw each other with a force proportional to the gravitational constant $G$.

This constant has characteristic physical dimensions, which is a sign of some phenomenon hidden in the mass interaction. All physicists know that the uniformity of this constant is not proven outside the solar system measurements. The astronomical observations of the solar system furnish very accurate figures of the product $(GM_i)$ for Sun and planets, but the masses are calculated in the *assumption* that $G$ is uniform.

In general relativity $G$ plays a very important role because its assumed universality compels the expanding cosmological models to predict (starting from a troublesome *big bang*) either a perpetual expansion at decreasing velocity, or a finite expansion with a final contraction at increasing speed (*big crunch*).

When in 1917 perceived this fact, Einstein introduced in the equations the so called cosmological constant which allowed a static model of the universe.

Considering the history of cosmology, we think that his choice was arbitrary, but not imprudent. As a matter of fact, the advanced cosmic observations presently compel the cosmologists, as properly recalled by L.Krauss (Ref.30), to reconsider the old cosmological constant[8]. This entitle us to think that the sagacity of Einstein guessed that expanding models (requiring the description of the *initial* universe) would have brought some turmoil about the basic concepts of physics.

## 2.1 – Critics of the concept of gravitational mass

Two orders of reasons require the concept of gravitational mass be disregarded.

The first one is the need of a «break» with the preceding theories of gravitation which did not recognise that a phenomenon involving thermodynamic systems (such as the gravitational masses) must satisfy the Second principle. Up to now, the gravitational theories avoided this fundamental test. This view is also supported in the work by H.Borzeszkowsky, T.Chrobok[9] which shows that the general-relativistic continuum thermodynamics is characterised by the absence of thermodynamical degree of freedom in the gravitational phenomena.

The only exception is the black hole thermodynamics introduced in the '70 by S. Hawking. In the present theory the energy degradation is introduced by the reduction of the quantum energy $E_i \rightarrow E_{i+1}$ in the collisions with particles.

The gravitational mass is currently considered the «source» (in analogy with the electric charges) of a field which generates perpetual motions of *macroscopic* bodies without *any* energy degradation, thus violating the Second principle.

The Newton's gravitational mass does not possess, according to the operational critics of science by P.W.Bridgman[10], the requisites of a field source. From the standpoint of epistemology it is not surprising that the gravitational mass gives rise to the embarrassing *unlimited* gravitational collapse.

The second reason regards the assumed uniformity of $G$ within the universe, which is not experienced. Conversely, several astronomers denounced, before and after the Hubble's discovery of the redshift, that between clusters of galaxies the gravitation appears to vanish[11,12].

In the present theory $G(r)$ descends from the equations of the quanta self-collision, which describe an exponential reduction in the intergalactic space (see Sec.2.4). Since $G(r)$ tends to zero on *large* scale, then the universe may be *static* without needing any cosmological constant.

Moreover, the figure of $G$ characterising the very dense stars may be much higher than the classical figure, in agreement with some recent experimental observations (see Sec.2.7).

### a) The origin of the mass

The cosmic quanta show an energy approximately $10^{-46}$ times the mass-energy of the electron. Their wavelength, equal to the Planck's length (see Sec.3.2), is much smaller than the electron. The cosmic quanta interact with any particle through a cross section $\sigma$ proportional to the mass, i.e. $(\sigma/m)$ is assumed to be constant for all particles.

This fundamental postulate gives the *mass* a unique basis. Classically, the mass was denounced by the inertia (and gravitation) of the bodies as well as of the light rays.

The energy of the electromagnetic field of charged particles leaded to attribute size and mass to the electric charges. But – in the words of R. Feynman[13] - this attempt leaded to «the failure of the classical electromagnetic theory [...], which is an unsatisfactory theory all by itself. There are difficulties associated with the ideas of Maxwell's theory, which are not solved by and not directly associated with quantum mechanics». In other words the physical basis of the mass is, independently of the quantum mechanical effects, not well established.

The paradigm of cosmic quanta gives this basis. The cosmic quanta must not be confused with the «gravitons» because they have no relation with this kind of particles devised to carry the old gravitational field. According to an open physical insight, the *field* is not indispensable in describing the physical phenomena. To this purpose the words of H.Bondi[14] may be recalled: «Since Maxwell, following Faraday, formulated electromagnetism as a field theory, the myth arose that all theories which proved successfully had to be field-theories. For this reason I considered a great step in physics the work of Wheeler and Feynman aimed to reformulating the Maxwell's theory in corpuscular terms».

*b) Some general implications*

Within the present paradigm the gravitational force ceases to be a *property* of the mass because it depends on the interaction of the *cosmic quanta* with the mass, as shown in the following paragraphs. Any variation of this gravitational force due to the varying mutual shielding transmits *obviously* with the velocity $c$ of the cosmic quanta. Consequently the axes of planetary orbits rotate as in general relativity (G.R.). The belief that the rotation of axes may be explained only with the curvature of the spacetime has been disproved by J.Montanus[15], which showed that "spacetime is flat everywhere, even in the presence of sources of gravitation". Within the tridimensional space, each mass undergoes the gravitational force through the net momentum released by two opposite beams of quanta: the uncollided beam and the beam which comes from the other mass after losing a little energy in a collision. This is the reason why the inertial-gravitational force may be described using the Newton's notation. The physical significance of the Newton's notations has been accurately analysed by H. Pfister in a recent paper[16], with particular reference to the concept of «free particle». In G.R. the gravitational interaction between masses depends on the modification of the *spacetime geometry*. According to a different thinking, the void space is filled by a high energy density which was estimated in 1967 by the theorist Y. Zel'dovic to be of about 55-120 orders of magnitude greater than the observed energy density of the universe[17].

Also the inertia of the mass comes from the interaction of particles with the cosmic flux. The main improvement introduced by the cosmic quanta is to give origin to *both* gravitation and inertia.

In the cosmic quanta paradigm the gravitational interaction is due to the modifications of the *energy-isotropy* of the cosmic flux induced by the collisions with matter (Compton effect).

The theory of inertia-gravitation here outlined differs from the newtonian and relativistic theories about the following items :

- the fundamentals of cosmology (the universe *evolves* without expansion)
- the strong gravity of the very dense masses (neutron stars and black holes)
- the absence of the unlimited gravitational collapse.

These differences are not surprising. After the season of the main contributions to the physics of XX century, Einstein proposed the general theory of relativity, devoting many years to the development of the mathematical formalism. As a matter of fact, he built an *observational* theory of the gravitational phenomena, which was particularly welcomed by the astronomers.

But the origin of the gravitation remained the old «gravitational» mass suggested by Newton three centuries ago. An intuitive concept which no longer fits the physical phenomena.

## 2.2 – The equivalence principle and the mechanics of special relativity

The equivalence principle, verified with an accuracy of 1 part on $10^{12}$ (Roll, Krotkov & Dicke, 1964), suggests that inertia and gravitation are two faces of a unique phenomenon.

Attempts to explain the inertia of bodies through the gravitational field of the distant masses of the universe (Mach's principle) did not go beyond the conceptual enunciation, owing to the fact that inertial forces manifest *instantaneously*. Hence the energy which generates them is *local* and has to be enormous since the inertial forces have no limit, according to the present knowledge.

Let's verify that the paradigm of the cosmic flux explains *both* gravitational and inertial forces.

### a) The relativistic mechanics

The principle of inertia states that a mass moving in the void, without external forces, continues to move indefinitely. A particle freely moving with velocity $\mathbf{v}$ within the space uniformly filled by quanta with energy $E_o = h_o \nu_o$, encounters in front quanta with an increased frequency $\nu_+ = \nu_o (1 + v/c)^{1/2} / (1 - v/c)^{1/2}$ and backwards a reduced frequency $\nu_- = \nu_o (1 - v/c)^{1/2} / (1 + v/c)^{1/2}$.

The relativistic Doppler effect is adopted to the aim of demonstrating that the cosmic flux paradigm fits special relativity. Let's consider two quanta colliding simultaneously on the particle from opposite directions with a net momentum

$$\Delta q = h_o (\nu_+ - \nu_-) / c = 2 h_o \nu_o \mathbf{v} / c^2 (1 - v^2/c^2)^{1/2}.$$

During the collision time of the waves $\tau_o = \lambda_o / c$, the flux $\phi_o$ determines on the particle with cross-section $\sigma$ a high number $\eta = \sigma \phi_o / 2\nu_o$ of double collisions (high coefficient of simultaneity).

The free particle collides with quanta showing a forwards energy density $\ni_+ = h_o \nu_+ \phi_+ / c$ which equals the backwards one $\ni_- = h_o \nu_- \phi_- / c$ owing to the conservation of the void initial energy density ($\ni_o = h_o \nu_o \phi_o / c$) during the collision with particles. These energy densities originate, forwards and backwards, the same pressure

(2.2.a) $$p_M = h_o \nu_+ \phi_+ / 4c = h_o \nu_- \phi_- / 4c = h_o \nu_o \phi_o / 4c$$

due to the forward reduction ($\phi_+$) and backwards increment ($\phi_-$) of the interacting flux. Incidentally, the isotropic pressure $p_M$ will result in Sec.2.5 the maximum physical pressure. Since no net momentum remains on the free particle ($\Delta \mathbf{v} = 0$), the total momentum of the simultaneously hitting quanta

(2.2.b) $$\mathbf{q} = \eta \Delta q = h_o \sigma \phi_o \mathbf{v} / c^2 (1 - v^2/c^2)^{1/2}$$

*coincides* with the momentum of the moving particle

(2.2.c) $$\mathbf{q} = m_o \mathbf{v} / (1 - v^2/c^2)^{1/2}$$

when the rest-mass is

(2.2.d) $$m_o = h_o \sigma \phi_o / c^2.$$

Putting $A_o = \sigma / m_o$ the constant ratio for any particle, the above equation implies also

$$h_o A_o \phi_o = c^2.$$

Resuming, the particle feels no force and moves freely in dynamical equilibrium with the cosmic flux, showing the relativistic mass-velocity dependence

$$m(v) = m_o / (1 - v^2/c^2)^{1/2}.$$

This means that special relativity agrees with a void structure *characterised* by a high flux of very small quanta. The eq.(2.2.d) also shows the equivalence between the particle mass-energy and the energy of the cosmic quanta which simultaneously interact

(2.2.e) $$m_o c^2 = E_o (\sigma \phi_o / \nu_o)$$

In high energy physics the mass of the generated particles (constituted of cosmic quanta) depends on the collision kinetic energy, i.e. the energy of the cosmic quanta that the inertial forces extract from the «void». Considering the high number of quanta which collide *simultaneously* $2\eta = \sigma \phi_o / \nu_o$ (for a nucleon $2\eta \approx 10^{50}$, see Sec.3.2), from eq.(2.2.e) one gets $E_o / mc^2 = (\nu_o / \sigma \phi_o) \approx 4 \times 10^{-51}$, i.e. the quantum energy $E_o$ results much lower than the mass-energy of a nucleon.

Deriving respect to time both sides of eq.(2.2.b) one correctly obtains the *two* inertial forces that the particle experiences due to the inertia of the cosmic flux

(2.2.f)    $\boldsymbol{F} = \mathrm{d}\boldsymbol{q}/\mathrm{d}t = m_o\,(\mathrm{d}\boldsymbol{v}/\mathrm{d}t)/(1 - v^2/c^2)^{1/2} + m_o\,\boldsymbol{v}\,v\,(\mathrm{d}v/\mathrm{d}t)/c^2(1 - v^2/c^2)^{3/2}$ .

 The first force opposes the variation of the speed *vector*. The second force is *tangential* and opposes the variation of the speed *modulus*. The tangential force makes sure that the angular momentum of an orbiting mass is not perfectly constant.

 In the classical theory, the velocity of the particle must be referred to the «inertial» frames, i.e. those anchored to the distant «fixed» masses of the universe.

 The cosmic flux, which fills the «bubble» of the universe where the galaxies are immersed, provides in principle an absolute reference system defining for each mass a velocity

 $v = c\,(v_*{}^2 - v_o{}^2)\,/\,(v_*{}^2 + v_o{}^2)$   depending on the forwards (or backwards) Doppler frequencies of the cosmic quanta.  This explains the preference that classical physics gave the so-called «inertial» systems.

### b) The undulatory-corpuscular nature of matter
 The dual nature of particles and waves was described by De Broglie defining the wavelength

$$\lambda^* \approx h\,/\,m_o\,v$$

of the dual wave  $h\nu^*$  having the same *momentum* of the particle    $m_o\,v \cong h\nu^*/\,c$ .

In the present paradigm each particle is characterised by a proper cross section in the collisions with cosmic quanta.  For the dual nature of matter, photons are characterised by a cross-section  $\sigma^*$  proportional to the equivalent mass  (basic postulate of this theory)

(2.2.g)    $\sigma^* = A_o\,m^*$ .

 The equivalent mass of a wave may be obtained writing the above equivalence of the momentum in the correct form

(2.2.h)    $m_o\,v\,/\,(1 - v^2/c^2)^{1/2} = h\nu/\,c = m^*\,v$ .

 In fact the very little rest-mass  $m_o$  of the photon becomes  $m^* = m_o\,/\,(1 - v^2/c^2)^{1/2}$   when the speed v  closely approaches the wave velocity $c$.  Thus one gets the mass-energy equivalence  $m^*\,c^2 = h\nu$  of a wave.  Substituting in eq.(2.2.g) one gets   $\sigma^* = A_o\,h\nu/\,c^2$ ,  which shows quantitatively that the corpuscular character (i.e. the cross-section) of waves becomes the more pronounced the higher is the photon frequency.

### c) An estimate of the constant $h_o$
 One has to question whether the constant  $h_o = E_o/v_o$  really equals the Planck constant related to the electromagnetic waves. Photons are complex objects with an energy $h\nu$ which, applying eq.(2.2.e) to the equivalent mass $m^*$ , results  $h\nu = E_o\,(\sigma^*\,\phi_o/v_o)$ , which generally is much greater than $E_o$ . Photons show whatever frequency, whereas the frequency of cosmic quanta is close to the fixed value $v_o$.  It seems that photons transmit a particular «state» of the constituent cosmic quanta. The ratio   $h\,/\,h_o$  may be obtained recalling eq.(2.2.d)

$$h\,/\,h_o = h\,A_o\phi_o\,/\,c^2.$$

Adopting the approximate figure $\phi_o \approx 10^{130}$  (see Sec. 3.2) one has   $h\,/\,h_o \approx 10^{69}$  which demonstrates the very different energy scale between cosmic quanta and electromagnetic photons.

## 2.3 -  The  newtonian gravitation: transparent masses
 An important feature of the interaction between quanta and *free*-particles is the flux *isotropy*.

 In fact the high coefficient of simultaneity of the cosmic quanta collisions determines an isotropic pressure (no net force) on the particle *without* energy variation of the incident quanta.

 The conservation of energy and momentum shows that some little energy is lost by a little fraction of the colliding cosmic quanta when the particle receives a real momentum  $\Delta(m\boldsymbol{v})$ , as it happens in the gravitational phenomena. In this case the equations of the Compton's effect hold.

From these premises, it is possible to demonstrate that between two masses there is a drawing force due to the balance of the total momentum of cosmic quanta which collide on each mass. As regards the «gravitational» effects, the masses can be considered constituted of nucleons (with mass $m$ ) adding the electron mass to that of the proton, because the very small momentum received by the electron is entirely transmitted to the atomic nucleus.

Putting $N_i = M_i / m$ the number of nucleons constituting the masses (at distance $r$ between the centres), the shielding angle the masses offer each other is given by

(2.3.a) $\qquad\qquad\qquad \gamma_j = \sigma N_j / 4\pi\, r^2$

when the mass diameter expressed in mean-free-path $\lambda_i = m / \sigma \delta_i$ (where $\delta_i$ is the mass density) is very small $D / \lambda_i \ll 1$, so the self-shielding of nucleons results negligible (transparent masses).

An evaluation of the constant $A_o = (\sigma/m)$ may be done comparing $\sigma$ to the cross-section of particles in the collision with photons (Compton effect). But the comparison is not correct because photons undergo *single* collisions (no simultaneity). Thus the major loss of energy takes place in the collisions with the electrons, a thing which does not happen to the flux of cosmic quanta. It appears necessary to find a figure of $\sigma$ which is self-consistent with the present theory.

Let's come back to the problem of the force measured between two spherical masses. It is sufficient to consider only the scant number of quanta which collide on a mass *after colliding* with the other mass.

Upon each nucleon of the mass $M_1$ we shall consider the quanta (comprised within the shielding angle $\gamma_2$ ) constituting couples of opposite direction formed by a quantum of momentum $E_o\, \boldsymbol{c}/c^2$ coming from space and a quantum with momentum $E_1 \boldsymbol{c}/c^2$ ($E_1 < E_o$) coming from a *preceding* collision on the mass $M_2$.

Whatever is the law of the angular diffusion in the quantum scattering with particles, this law will be the same both in the collision $E_o \rightarrow \psi(\theta) E_1(\theta)$ and in the collision $E_1 \rightarrow \psi(\theta) E_2(\theta)$. Due to the small energy $E_o$ respect to the particle mass-energy, we assume the simple distribution law $\psi(\theta) \cong 1$.

Hence the first quantum gives a net momentum $\Delta q_1$ which tends to *get closer* the masses

(2.3.b) $\qquad\qquad\qquad \Delta q_1 = E_o / c - \int_o^\pi (E_1(\theta)/c)\cos\theta \sin\theta \; d\theta$

whereas the second quantum gives a net momentum $\Delta q_2$ which tends to *remove* the masses

(2.3.b) $\qquad\qquad\qquad \Delta q_2 = E_{1m}/c - \int_o^\pi (E_2(\theta)/c)\cos\theta \sin\theta \; d\theta$

where $E_{1m} = (1 - K_o)E_o$ (being $K_o = E_o / mc^2 \ll 1$) is the average of the quanta $E_1(\theta)$ diffused by each particle of $M_2$ toward the mass $M_1$.

The energy of a quantum after the $i$-th collision (Compton effect) is

(2.3.c) $\qquad\qquad\qquad E_i(\theta) = E_{i-1} / [1 + E_{i-1}(1 - \cos\theta)/ mc^2].$

Substituting in the eq.(2.3.b) one obtains

$\qquad\qquad\qquad\qquad \Delta q_1 = E_o / c - 2\, K_o\, E_o / 3c$

and analogously $\quad \Delta q_2 = E_{1m}/c - 2\, K_{1m} E_{1m}/ 3c$ , where $K_{1m} = E_{1m}/mc^2 \ll 1$.

From the balance of the net momentum left on the mass $M_1$ by the flux comprised within the shielding angle $\gamma_2 = \sigma M_2 / m\, 4\pi\, r^2$, a force takes place pushing the masses between them

(2.3.d) $\qquad\qquad\qquad F(r) = (\Delta q_1 - \Delta q_2)\,(\sigma N_1)\,(\phi_o\, \gamma_2)$

which by substitution becomes

(2.3.d) $\qquad\qquad\qquad F(r) \cong [(E_o - E_{1m})\, A_o^2\, \phi_o / 4\pi\, c]\, M_1\, M_2 / r^2$ .

This force is due to the radiation pressure of the cosmic quanta, but for simplicity we continue to name it the «gravitational» force. Substituting $E_{1m}$ one gets finally

$$F(r) = [K_o E_o A_o^2 \phi_o / 4\pi c] M_1 M_2 / r^2 .$$

The quantity in brackets appears to be statistically constant and, for appropriate figures of the unknown constants, it becomes the gravitational constant *measured* in the laboratories

(2.3.e) $\qquad\qquad\qquad K_o E_o \phi_o A_o^2 / 4\pi c = G.$

To this point we have to verify that the derivation of the force $F$ is correct even from the standpoint of the Second principle, thus pertaining to the dissipative phenomena.

The proof is given by the eq.(2.3.d) which defines the force proportional to $(E_o - E_{1m})$, that is to the energy given up by each quantum.

It can be useful to recall some experimental facts to give a more physical description of the genesis of the gravitational force. The first measurement of $G$ was made by Cavendish in 1798 obtaining $6.754 \times 10^{-11}$ Nm$^2$/kg$^2$. As reported by A.H.Cook [18], in the two successive centuries many other measurements were performed with different methods by Poynting (1891), Boys (1895), Braun (1897), Heyl (1930), Zahradnicek (1933, obtaining $G = 6.659$), Heyl and Chrzanowski (1942), Luther & Towler (1982, which obtained $G = 6.6726$) up to the most recent (2001) by Quinn, Speake et al.[19], which found $G = 6.67559$.

Substantially the great development of technologies, which allowed to measure other physical constants with a reliability better than $10^{-6} \div 10^{-7}$, still gives values of $G$ which differ in the fourth figure. Obviously, we have to consider that the measurement of a small force is sensible to many disturbs. Notwithstanding , it is possible to put forward the hypothesis that the persistent discrepancies between various measurements might by due to the complex genesis of this force.
This genesis may be understood on simple physical basis recalling, for instance, the problem of the air convection currents between the spheres.
Since these currents may alter the result, in all experiments the torsion balance was contained within a box. To the aim of eliminating the trouble, Braun (1897) tried to reduce the pressure in the ampoule, without observing any practical effect. In fact the calm air at atmospheric pressure does not create appreciable disturbance to the zero-balance of the momentum that hitting air molecules give isotropically to the spheres. Considering carefully this fact , it appears that the limited void capacities of the Braun's epoch prevented him to make an observation which is not banal. Reducing the air pressure within the glass up to the point when the mean-free-path of the molecules is comparable to the minimum distance between the sphere surfaces, he could have noticed an enormous increase of the measured force due simply to the very small air *depression* between the spheres. Obviously this phenomenon vanishes when all molecules are eliminated from the ampoule.
With a thought experiment, let's substitute the incident air molecules with the penetrating cosmic quanta. Then a «gravitational» force will appears , because each mass sends to the other a (proportional) number of quanta with momentum $E_1/c < E_o/c$.

## 2.4 - The generalised gravitational force

The high density celestial bodies (white dwarfs, neutron stars, ecc.) can be treated with a gravitational equation similar to eq.(2.3.d) when we take into account the fact that quanta coming from these stellar masses have undergone numerous collisions within them. Any exiting quantum undergoes a number of collisions which is very complicate to calculate exactly, because there are both quanta which crossed the whole mass and quanta reflected after a small path. From the energetic standpoint the spectrum of the leaving quanta may be substituted, in a elementary approach, by the average number of collisions $n$. The optical thickness of a mass (radius $R_i$, average density $\delta_i$ ) can be also introduced

(2.4.a) $\qquad a \cong (4/3) R_i \, \delta_i A_o = M_i A_o / \pi \, R_i^2.$

From considerations of energy conservation, a simple approximate expression has been found between the number of collisions and the optical thickness

(2.4.b) $\qquad n \cong a / (1 - e^{-a}).$

which gives obviously $n = 1$ for transparent masses ($a \ll 1$). This relationship is accurate for low density celestial bodies, which obey the classical gravitation (see Sec.2.6).

### a) *The gravitation of the opaque masses*

After many collisions in the mass, the leaving quanta show a reduced energy $E_n$ which can be obtained reiterating eq.(2.3.c):

(2.4.c) $\qquad E_n = E_o / (1 + n \, E_o / mc^2) = E_o / (1 + n K_o).$

Let's see the figures taken by the number $n$. In the case of planets $n$ is around some unity (e.g. Earth, optical thickness 2.2 , $n \approx 2.47$), while for the stars it goes up to hundreds (e.g. Sun, $n \cong a \approx$ 62). In the case of the white dwarfs $n$ goes up to $10^5 \div 10^7$.

When a transparent mass $M$ interacts with an opaque body (mass $M_i$, radius $R_i$) the gravitational force is linked to the shield of the opaque mass, which emits weak quanta of energy $E_n$. The fraction of the flux $\phi_o$ which collides with the mass $M$ after colliding with the opaque mass $M_i$ is

$$\gamma_i \cong (\pi R_i^2 / 4\pi \, r^2) \, (1 - e^{-a}).$$

For continuity this attenuation gives, when the optical thickness $a \ll 1$, the shielding angle previously seen for the transparent masses $\gamma = A_o \, M / 4\pi \, r^2.$

Let's now calculate the gravitational force $F$ between the transparent mass $M$ and a large mass $M_i$ with the attenuation $\gamma_i$. To this aim we write the balance between the momentum given to the mass $M$ by quanta coming from space $\Delta q_1 = E_o / c$ and by quanta coming from the opaque mass $\Delta q_n = E_n / c$

(2.4.d) $\qquad F = (\Delta q_1 - \Delta q_n) \, (\sigma M/m) \, \phi_o \, \gamma_i = (E_o - E_n) \phi_o A_o \, M \, (R_i^2 / 4 c r^2) \, (1 - e^{-a}).$

An interesting application of this equation is the calculation of the gravitational force between an object on the Earth and the Earth itself. In this case the acceleration of gravity is $g = F/M = 9{,}81$.

Considering the real path through the planet of the colliding quanta, we approximately obtained[20] a figure $A_o \approx 4.7 \times 10^{-11}$ together with a figure of the Earth optical thickness $a \approx 2.2$.

A more accurate value of $A_o$ will be found in Sec.3.2 through the dimensional analysis.

Substituting in eq.(2.4.d) the identity $\pi R_i^2 = A_o M_i / a_i$ and taking into account eq.(2.4.c), one gets the expression

(2.4.e) $\qquad F = [n /(1 + n K_o)] \, [(1 - e^{-a})/a ] \, G \, M_i \, M / r^2$

where the gravitational constant is multiplied by the coefficient (being $n K_o \ll 1$ for all stars)

(2.4.f) $\qquad \Psi = (n / a) \, (1 - e^{-a})$

which contains the ratio between the mean number of collisions and the optical thickness.

When eq.(2.4.b) is verified (e.g. for low density stars) one gets immediately $\Psi \cong 1$, so eq.(2.4.e) takes the familiar form of the classical Newton's law.

However a remarkable difference arise in the case of dense stars (for which $e^{-a} = 0$) since the ratio ($n /a$) becomes greater than 1. This happens because the quanta undergo a small deviation in the collisions, so the trajectory bends and they undergo within the mass an average number of collisions $n$ greater than $a$, which equals the number of collisions if quanta were moving in a straight line.

The gravitational equation in the case of neutron stars becomes simply

(2.4.g) $\qquad F \cong (n/a) \, G \, M_i \, M / r^2.$

In Sec. 2.6-2.7 the factor ($n/a$) will be evaluated by doing the analysis of the star stability.

### b) *The mean free path of cosmic quanta*

It seems incredible that such small objects as cosmic quanta denounce a mean free path less than photons, which seem to undergo no collisions between them within the observed size of the universe. This depends on the very small flux $\phi_f$ of photons compared with the cosmic flux.

The photon m.f.p. $l_f = c / \sigma_\pi \phi_f$ depends on the equivalent cross-section $\sigma_\pi = A_o h \nu / c^2$ given by eq.(2.2.g). Since the flux of photons is $\phi_f = W / h \nu$, the m.f.p. becomes

(2.4.h) $$l_f = c^3 / A_o W$$

where the flux of electromagnetic power $W$ (summing the CMB power and the electromagnetic emissions by stars in the universe) is about $W \approx 10^{-5} \div 10^{-4}$ w/m².

The resulting $l_f \approx 10^{40}$ takes figures well beyond the observed radius of the universe.

Viceversa the cosmic quanta m.f.p. $l_o = c / \sigma_o \phi_o$ depends on the very high flux $\phi_o \approx 2.43 \times 10^{130}$ (Sec.3.2) multiplied by the extremely small quantum cross section $\sigma_o$ whose value can be assumed from dimensional analysis which indicates a small area $s_o = l_P^2 (2\varepsilon_o G m^2/e^2)^2 = 2.66 \times 10^{-143}$ related to the Planck's length $l_P$ multiplied by the ratio (gravitational force / Coulomb's force) between two nucleons. Putting $\sigma_o = s_o$ in eq.(2.4.i), one has

(2.4.i) $$l_o = 4.62 \times 10^{20}.$$

### c) The gravitational force between very distant masses

When the distance between two galactic masses is comparable with $l_o$, a beam $\Psi_s$ of quanta $E_o$ arrives on the first mass flying together with the beam of quanta $E_1$ scattered by the second mass. This beam $\Psi_s$ is due to the self-collisions between the cosmic quanta $E_o$ in the space within the shielding angle $\gamma(r)$. The effect of the self-collision becomes sensible only when the distance of the masses is of the order of the intergalactic distance. In this case, putting $\Sigma_o = 1 / l_o$ the macroscopic self-cross section of quanta, the gravitational force between the galactic mass $M_c$ [which requires the average star gravity factor $\underline{(n/a)}$ given by eq.2.7.a] and a very distant mass $M$ is

(2.4.l) $$F = \sigma \underline{(n/a)} (M_c/m) \gamma(r) \phi_o [ E_o - E_o\!\int_o^r \underline{\Sigma} \exp(-\underline{\Sigma r}') \, dr' - E_1 \exp(-\underline{\Sigma r}) ]/c$$

where $\underline{\Sigma r} = \int_o^r \underline{\Sigma}(r') dr'$ is the optical distance and $\underline{\Sigma}(r) = [ l_o^{-1} + A_o \delta_u(r) ]$ is the total cross-section encountered by the cosmic quanta in the intergalactic space, $\delta_u(r)$ being the gas density. Rearranging and substituting $\gamma(r) = (M/m)\sigma/2\pi r^2$, one gets

$$F = (\sigma/m)^2 \underline{(n/a)} M_c M \phi_o (E_o - E_1) \exp(-\underline{\Sigma r}) / 2\pi c r^2$$

which, recalling eq.(2.3.d), becomes the force between very distant masses

(2.4.m) $$F = (n/a) G \exp(-\underline{\Sigma r}) M_c M / r^2.$$

As long as the density of the intergalactic matter $\delta_u(r) << (A_o l_o)^{-1} \cong 5 \times 10^{-11}$, the optical distance becomes $\underline{\Sigma r} \cong r/l_o$ and the varying gravitational constant between galaxies results

(2.4.n) $$G(r) \cong \underline{(n/a)} G \, e^{-r/l_o}.$$

Since the average distance between clusters of galaxies is about 100 times the m.f.p. $l_o$, it follows that clusters are not gravitationally bound.

## 2.5 - The paradox of the unlimited gravitational collapse

Between the known celestial bodies, the most dense are the neutron stars. They are objects with a mass greater than 3-4 solar masses and a radius $R$ around a few ten kilometres, characterised by rapid rotations (pulsar) with periods from some milliseconds up to some seconds. Their gravitational acceleration is high, otherwise the enormous centrifugal force would remove layers of matter from the equator.

The mean number of collisions $n$ for neutron stars is much greater than the optical thickness $a$, which is of the order of $10^{10} \div 6 \times 10^{11}$. The density of the neutron stars may reach about $10^{16} \div 10^{18}$ kg/m³. During the contraction of neutron stars, the internal pressure increases in such a way that no organisation of the elementary particles is able to sustain the pressure due to the gravitational mass. This fact allows the embarrassing *unlimited* gravitational collapse.

Conversely, within the present paradigm the stars cannot collapse beyond a certain density. In the classical paradigm the internal pressure is originated by the force on each particle $f_{cl} = GM\,m/r^2$ due to the gravitational mass .

In the cosmic quanta paradigm, the force on each peripheral particle is due to the net radiation pressure produced by collisions of the inward quanta ($E_o/c$) and the outward quanta ($E_n/c$). Therefore the mass is compressed by the outside cosmic radiation pressure and consequently the supernova explosion originates necessarily *after* the star implosion/collapse.

The force on peripheral particles of an opaque star ($e^{-a} \cong 0$) depends on the number $n$ of collisions suffered by the outgoing quanta

(2.5.a) $\qquad\qquad f = \sigma\,\phi_o\,(E_o - E_n)/\,4c = \sigma n K_o\,\phi_o\,E_o\,/\,4c\;(1 + nK_o) = \sigma\,p(n)$

where $p(n) \cong nK_o\phi_oE_o/4c$ is the cosmic radiation pressure (being $nK_o \ll 1$ for any star).

Recalling eq.(2.3.e) this peripheral force becomes

$$f \cong \pi\,n\,G\,m\,/A_o$$

whereas at the distance $r$ from the centre it assumes the form

(2.5.b) $\qquad\qquad f(r) \cong \pi\,n(r)\,G\,m\,/A_o$

where $n(0) = 0$, since $f(r)$ vanishes by reason of central symmetry.

The thermodynamic pressure of the gas $p_o(\underline{r})$ at a distance $\underline{r}$ originates by summing the forces $f(r)$ acting on each particle placed in the shell of mass $\Delta M(\underline{r})$ between $R$ and $\underline{r}$

$$p_o(\underline{r}) \cong (1/4\pi\,\underline{r}^2) \int_{\underline{r}}^{R} 4\pi\,r^2\delta(r)\,f(r)\mathrm{d}r\,/m = \underline{n}\,G\Delta M(\underline{r})\,/4\,A_o\,\underline{r}^2\;.$$

Since the central pressure is finite, the ratio $n(\underline{r})/\,\underline{r}^2$ is everywhere finite and takes the figure $n/R^2$ at the periphery. Considering the radius $\underline{r}$ which halves the mass $M$, the condition $\underline{n}\,/\,\underline{r}^2 \cong n\,/R^2$ is likely verified. The average gravitational pressure $p_o$ takes place when $\Delta M(\underline{r}) = M/2$, thus the preceding equation gives simply

(2.5.c) $\qquad\qquad p_o \cong n\,GM/8\,A_oR^2 \cong 0.324\,n\,G\delta^{2/3}M^{1/3}/A_o$ .

From this equation it is possible to evaluate $n$ for white dwarfs and neutron stars imposing that $p_o$ balances the average pressure of the degenerate matter $p_d = \pi^2\hbar^2\delta^{5/3}/\,5m_e\,m^{5/3}$ plus the pressure of the electromagnetic radiation.

Thus the stability condition of the star, namely $p_o \cong p_d + p_r$, requires that

(2.5.d) $\qquad\qquad n_{eq} \cong 2.26\mathrm{x}10^7\,\delta(1+\chi)\,/M^{1/3}$

where $\chi$ is the ratio $p_r/p_d$. When the density does not exceed the figure of the white dwarfs, the radiation pressure may be neglected, so the above expression gives $n_{eq} \approx 10^5 \div 10^6$.

From the theoretical standpoint, eq.(2.5.a) shows that if hypothetically the collapsing mass reached the exceptional condition $nK_o > 1$, the cosmic radiation pressure $p(n)$ upon particles would approache the maximum theoretical figure $p_M = \phi_o\,E_o\,/4c$.

Actually the degenerate matter cannot pressurise up to $p_M$ or contract beyond a finite density $\delta_x$, thus excluding the unlimited collapse. At a first approximation, the maximum density $\delta_x$ equals the density of nucleons very close each other, i.e. $\delta_x \approx 10^{27} \div 10^{28}$ which is related to the ratio $m/\sigma^{3/2}$.

Recalling that $\pi\,G = 2\,p_M\,A_o\,K_o$ and substituting in eq.(2.5.c), the maximum gas pressure can be put in the form $p_{ox} = p_M\,(n_x/\,3)\,\delta_x R\,A_oK_o$ which gives the maximum figure of the ratio $(p_{ox}/p_M) = 4.55\mathrm{x}10^{-8}(1+\chi)\,\delta_x^{5/3}A_oK_o$ . Assuming the figure of $K_o$ calculated in Sec.3.2 , the ratio $(p_{ox}/p_M)$ results of the order of $10^{-8}$

## 2.6 - The limit of accuracy of the classical gravitation

Comparing the classical gravitational force $f_{cl} = GMm/R^2$ upon a peripheral particle of an opaque star with the cosmic pressure force $f \cong \pi\, n\, G\, m/A_o$, the resulting relationship $M/R^2 \cong \pi\, n/A_o$ states that when the number of collisions equals the optical thickness

$$(2.6.a) \qquad\qquad n \cong A_o M/\pi R^2 = a$$

the Newton's gravitation is accurate *not only* for transparent masses (see eq.2.4.b), but even for *opaque* masses $n \cong a$, at least up to a certain limit of $a$.

Let's try to calculate the number of collisions of quanta outgoing from celestial bodies mostly constituted by a gas of atoms (mass $m_n$), such as the ordinary stars.

The stability of these bodies against collapse is assured as long as the gas pressure $p \cong \delta(k/m_n)T$ equals the average gravitational pressure $p_o \cong n\,(1-e^{-a})\,GM/8A_o R^2$ which considers also the very low density stars appearing in the HR diagram. Therefore, the average number of collisions which assures the stability ($p \cong p_o$) of the ordinary stars results

$$(2.6.b) \qquad\qquad n_{eq} \cong 8\delta(k/m_n)T A_o R^2/(1-e^{-a})GM.$$

Considering the average Sun parameters, one gets $n \cong 60 \div 70$, in agreement with the calculated optical thickness ($a = A_o M/\pi R^2 \cong 62$). The same happens for other stars of the main sequence in the HR diagram. This result permit us to explicit the gravity factor $\Psi$ appearing in the generalised equation (2.4.e) of the gravitational force

$$(2.6.c) \qquad\qquad \Psi = (1-e^{-a})(n_{eq}/a) \cong 6\,(k/m_n)T R/GM.$$

This factor is close to *unity* for the ordinary stars, proving accurate the Newton's gravitation.

## 2.7 - The gravitation of superdense stars

The factor $(n_{eq}/a)$ notably increases when we consider the pressure of the degenerate matter within a white dwarf or a neutron star. Recalling the number $n_{eq}$ of collisions at stability given by eq.(2.5.d), the gravity factor is defined by

$$(2.7.a) \qquad \Psi = (n_{eq}/a) \cong 3.6\text{x}10^{17}(1+\chi)/R\,M^{1/3} \cong 5.8\text{x}10^{17}\delta^{1/3}(1+\chi)/M^{2/3}.$$

For white dwarfs it may be somewhat grater than 1. Substituting the typical parameters of the neutron stars, the above expression gives figures around $9\text{x}10^2(1+\chi)$, which increases up to $2\text{x}10^6(1+\chi)$ when the density of a collapsed neutron star equals the maximum density $\delta_x \cong 10^{27}\div 10^{28}$ suggested in Sec.2.5. In these conditions the redshift is such that the star becomes invisible (black hole). Let's recall that here the term "black hole" describes an invisible neutron star, without any reference to the concept of Schwarzschild's radius (see Sec.2.8). To mark the difference we shall use in advance the term "obscure object". Considering that in the neutron stars the radiation pressure may overcome the gas pressure ($\chi>1$), the gravitational force of a wholly collapsed neutron star may be more than 6 orders of magnitude greater than the Newton's force.

An interesting result has been recently found by Miller et al.[21] using data by the X-ray Newton satellite (ESA) which accidentally observed in a galaxy distant 170 million light years the effects of a black hole «swallowing» the nearest bodies. The gravitational mass, calculated through the observation of three luminous orbiting bodies with 27 hours period, resulted about 300,000 times the Sun mass. The «swallow» of orbiting bodies is unknown to the classical gravitation because the constancy of $G$ maintains any celestial mass in a fixed orbit.

On the contrary, eq.(2.7.a) shows that the effective gravity $G(R) = (n_{eq}/a)G$ of a dense star depends on its radius. Hence $G(R)$ increases when $R$ reduces and the orbiting mass really flows towards the dense body. The «swallow effect» does not happen when the dense body is *not* collapsing.

Previously an other valuable result was obtained observing in the centre of our Galaxy, near Sagittarius A*, a huge invisible mass, which recently resulted equal to $3.7\text{x}10^6$ Sun masses after a ten-years observation of an orbiting star with a period of 15.2 years[22]. These huge masses, which are abnormal respect to the progenitor stars, are currently explained assuming that the mechanism of mass accretion ascertained in the study of binary stars, works even on the galactic scale.

D.Figer found[23] that in the Arches clusters there is no star heavier than 150 solar masses. The probability that observations agree with the absence of upper limit equals $10^{-8}$.

Probably the observed black holes do not have such huge masses. In the present theory the observed *huge gravity* does not correspond to a high gravitational mass $M^*$, but to the effective gravity $G(R) = (n_{eq}/a)G$ originating from the weakened cosmic quanta $E_n$ leaving the black hole. The observed mass $M^* = (n_{eq}/a)M$ depends on the black hole real mass $M$ coherent with the progenitors.

In general the gravity factor $(n_{eq}/a)$ seems essential in assessing the correct mass of very dense stars. For instance there is evidence, according to A.Aguirre et al.[24], of some «lack» of galactic mass since the observed star rotation velocities are practically constant along the galaxy radius, contrarily to the decreasing Newton's velocities $v^2(r) = GM(r)/r$. This may be explained introducing the gravity factor of the dense stars, provided their number is high enough. For instance, Kroupa[25] thinks that the most abundant population in galaxies is the (red, white, brown) dwarf stars, which have a gravity factor somewhat greater than unity. This might account for some "missing" mass. But it's not easy to evaluate the fraction of white dwarfs or neutron stars.

## 2.8 – The gravitational redshift

In astrophysics the «black holes», originating from the collapse of neutron stars, are bodies whose escape velocity equals the velocity of light. It is interesting to evaluate the braking force of the cosmic quanta which reduces the energy of the emitted photons.

The photons interact through the cross section $\sigma^* = A_o m^*$ (see eq.2.2.h) with the quanta $E_o$ coming from space and with the weak quanta $E_n < E_o$ coming from the black hole. From eq.(2.4.g) the gravitational force on the photon is

(2.8.a) $$f \cong (n_{eq}/a) \, G \, M_i \, m^*/r^2$$

where $m^* = h\nu/c^2$ is the equivalent mass and the ratio $(n_{eq}/a)$ is evaluated with the procedures shown in Sec.2.6 and Sec.2.7. The reduction of the photon energy satisfies the equation

(2.8.b) $$d(h\nu)/dr = -f$$

which gives the gravitational redshift of the mass $M$ observed at large distances

(2.8.c) $$z_g = (\nu_o - \nu)/\nu = \exp[(n_{eq}/a)GM/c^2R] - 1 = e^\alpha - 1$$

As a consequence only a negligible fraction of the emitted electromagnetic radiation escapes from a neutron star when (recalling eq.2.7.a) the exponent $\alpha \cong 2.67 \times 10^{-10} M^{2/3}/R^2 \cong 6.91 \times 10^{-10} \delta^{2/3}$ takes figures equal to a few ten. For instance when the observed radiation is reduced of $z_g \cong e^\alpha \approx 10^{10} \div 10^{12}$ times, the neutron star becomes invisible on the whole spectrum (black hole).

In General relativity the blueshift of a radiation $\nu_o$ *arriving* on a dense star is given by the exact formula $z_{blu} = (\nu - \nu_o)/\nu = 1 - (1 - 2GM/c^2R)^{1/2}$. Assuming the energetic effects of gravity do not depend on the direction of the photon, the radiation *leaving* a dense neutron star shows a redshift $z_g \approx 1/(1 - 2GM/c^2R)^{1/2}$ which loses any physical significance when $R$ goes beyond the Schwarzschild radius. Conversely, eq.(2.8.c) shows a continuous drop of the observed radiation frequency in accord with the thermodynamic requirements for macroscopic bodies.

The new paradigm of cosmic quanta gives a conceptual improvement respect to the classical field theory, because it explains without difficulties the permanence of the gravitational force (due to the weak cosmic quanta escaping from the dense mass) *outside* the black hole.

On the contrary, the classical theory does not explain why the field-carrying waves don't undergo any redshift, whereas the photons do. In General relativity this problem is transferred to the bending of the spacetime, whose mathematical formalism now appears to express the *physical* actions transmitted by the cosmic flux.

A recent work by O.Dreyer et al. [26] proposes a test for establishing the possible difference between the spectra of the observed radiation arising from relativistic black holes and other sources (e.g. neutron stars, strange matter, boson stars). This proposal may be useful also in clarifying the above problem.

## 3 – THE ORIGIN OF THE NUCLEAR FORCES

The degenerate matter of the observed neutron stars shows average densities around $10^{13} \div 10^{16}$.

From the outer layer ($\delta = 10^8 \div 10^9$) constituted by ionised nuclei and electrons (like white dwarfs) the density rises up to the figure $4 \times 10^{14}$ at which single neutrons begin to detach from the nuclei. Within the core all matter is constituted of superfluid neutrons, whose density approaches that of the atomic nuclei. The inward force which confines the peripheral particle (cross section $\sigma$) of a star is

(3.1.a) $\qquad f(n,r) = \sigma \gamma(r) \, \phi_o (E_o - E_n) / 4c \cong \sigma n K_o \, \gamma(r) \, p_M$

where $n$ is the number of collisions within the star , $p_M$ is the maximum cosmic radiation pressure, $\gamma(r)$ is the shielding angle of the star mass.

Recalling eq.(2.4.g), the force becomes

$$f(n,r) = (n_{eq} \, /a) GMm \, / r^2.$$

This force shows a very large range of figures, since $n$ varies from $1$ up to $\approx 10^{24}$.

It may be useful to resume the various ranges of $n$ showing the various features of the force due to the cosmic pressure.

For any star density, eq.(2.5.d) gives the corresponding number of collisions $n_{eq}$ .

- $\quad 1 < n_{eq} < 10^5 \quad \Rightarrow \quad f(n,r) = GMm / r^2$ is the classical Newton's force ($n = a$)
- $\quad 10^5 < n_{eq} < 10^{13} \quad \Rightarrow \quad f(n,r)$ with $(n_{eq} \, /a)$ rising to about $9 \times 10^2 (1+\chi)$ for neutron stars
- $\quad 10^{13} < n_{eq} < 10^{24} \quad \Rightarrow \quad f(n,r)$ with $(n_{eq} \, /a)$ rising up to $2 \times 10^6 (1+\chi)$ for collapsed black holes.

In particular considering the maximum theoretical density $\delta_x \approx 10^{28}$ (see Sec.2.5), the drawing force $f(n,r)$ may rise to $2 \times 10^6 (1+\chi)$ times the Newton's force.

To this point a question arises: is it conceivable that the drawing force due to the cosmic pressure is capable of confining protons within the atomic nucleus ?

The usual stable nuclei made by a few hundred nucleons are structures *transparent* to the cosmic quanta. For instance the optical thickness of the heaviest stable nucleus (Uranium) takes a figure $a_N = (4/3) \delta_N r_N A_o \approx 10^{-5}$. Since the quanta incident on a transparent structure do *only one* collision, then the cosmic pressure acts upon nuclei in a different way.

### 3.1 - The action of the cosmic radiation pressure upon nucleons

To express concisely the situation about the theory of nuclear forces, R. Feynman[13] argued: «There are other forces – like the nuclear forces – that have their own field theories, although no one knows whether the current theories are right». For instance, the classical theory of the intermediate particles (mesons, 1934) comes from applying to them the equation governing the field-carrying particles of electromagnetism (photons). To build a «nuclear» field it was necessary to *hypothesise* something similar to the electric charge, the so called «nuclear charge», which generates only drawing forces analogously to the «gravitational» mass. Unfortunately, the last one was unable to explain the gravitational field by means of field-carrying particles.

The current physical theory of the nuclear forces[27] points out a series of rules assessing all known characteristics about nuclei.  The following few rules describe the force which confines the nucleons:

- the nuclear force between nucleons does not depend on the electric charge
- each nucleon interacts with a limited number of nucleons in close proximity
- the short range nuclear forces practically vanish outside the nucleus
- the nuclear forces are not of central type.

### a) The drawing force between two nucleons

The cosmic pressure acts upon elementary particles in a different manner respect to the macroscopic bodies, as described in Sec. 2.3. When we consider two very close particles, the quanta $E_{i+1}(\theta)$ produced by the collisions of the quanta $E_i$ show an unknown angular dependence due to the law of scattering of that particle. Let's consider two nucleons at distance $r$ with mutual shielding angle

$$\gamma(r) = \sigma/2\pi\, r^2$$

and apply to each particle the energy conservation between the *arriving* (left side) and the *leaving* (right side) quanta, taking also into account that the cosmic quanta, arriving within the shielding angle from space $(E_o)$ and from the other nucleon $(E_1)$, give the matter an energy  $E_o - E_1 = K_o E_1$  at each collision.  The quanta arriving *outside* the shielding angle $\gamma(r)$ do not lose energy because the particle undergoes a (dissipating) force *only* along the direction of the force.

The  (non-isotropic) scattered quanta  $\sigma\psi_1(\theta)E_1(\theta)$ gives rise to a beam , with energy  $\sigma\gamma E_1\psi_1$, which collides with the other nucleon. This beam produces the scattered flux  $\phi_2$, a beam of which, with energy  $\sigma\gamma E_2\psi_2$,  produces on the other nucleon the scattered flux $\phi_3$ and so on.  Resuming , the principal equations of energy conservation, before and after each collision, are

(3.1.a)  $$\phi_o E_o\,(1-\gamma/2)\ = E_1\phi_1 + \gamma\,K_o E_1\phi_o$$
$$\gamma E_1\psi_1 = E_2\phi_2 \quad ; \quad \gamma E_2\psi_2 = E_3\phi_3 \quad ; \quad \gamma E_3\psi_3 = E_4\phi_4 \quad ; \quad \ldots$$

where each scattered flux is   $\phi_i = \int_o^\pi \psi_i(\theta)\sin\theta\;d\theta \quad (i = 1, 2, 3 \ldots)$.

The energy  $(E_o - E_1) = K_o E_o$  lost by quanta produces (see Sec.2.3) the gravitational force.  In fact the momentum associated with the energy   $\gamma\,(E_o - E_1)\,\phi_o$  of two opposite incident quanta gives rise to the force

$$f_G(r) = \sigma\gamma(r)\,(E_o - E_1)\,\phi_o\,/c = \sigma^2 K_o E_o\phi_o\,/2\pi c r^2$$

which, compared with eq.(2.3.d), appears to be the gravitational force   $f_G(r) = G\,m^2/r^2$.

Since  $K_o \cong 10^{-50}$  (see Sec.3.2) is very little compared with the angle $\gamma(r)$  (which is not less than about $10^{-17}$ near the nucleus), the difference  $(E_i - E_{i-1})$  may be everywhere neglected in eqs.(3.1.a) which become

(3.1.b)  $$\phi_o\,(1-\gamma/2)\ = \phi_1$$
$$\psi_1\,\gamma\ = \phi_2 \quad ; \quad \psi_2\,\gamma = \phi_3 \quad ; \quad \psi_3\,\gamma = \phi_4 \quad ; \quad \ldots$$

Now let's do the balance of *momentum* between the arriving and the leaving quanta.  It is sufficient to consider only the quanta incident within the shielding angle, because outside $\gamma(r)$ the flux is isotropic and does not gives rise to any differential momentum. Since the shielding angle between nucleons shows normally within nucleus a little figure $\gamma(r) \approx 10^{-8}$, the quanta that each nucleon receives from the other can be treated as beams. In particular the beams give the particle the total momentum  $\gamma(E_1\psi_1/c + E_2\psi_2/c + E_3\psi_3/c \cdots)$  contrary to the momentum   $\gamma E_o\phi_o/c$  of the beam coming from space which hits the opposite face of the particle.

From this balance the following drawing force arises

(3.1.c) $$f_N(r) \cong \sigma \gamma(r) E_o \, [\phi_o - \textstyle\sum_n \psi_n] / c$$

where the single beams result $\psi_n = \gamma^{n-1} \, \psi_1$ $(n = 1, 2, 3, \ldots)$.

Due to the lack of quanta $E_o$ within the shielding angle, the flux $\psi_1(\theta)$scattered by the particle results non-isotropic, so that it can be approximated by

$$\psi_1(\theta) \cong \phi_1 \, (1 - \varepsilon \cos\theta).$$

It shows a minimum along the direction $(\theta \cong 0)$ facing the other nucleon, so the first beam results to be $\psi_1 = \psi_1(0) = \phi_1(1-\varepsilon)$. Along the opposite direction $(\theta \cong \pi)$ the arriving flux $\phi_o$ does not suffer reduction (i.e. $\psi_1(\pi) = \phi_o$) and is approximated by $\psi_1(\pi) \cong \phi_1(1+\varepsilon)$. This means that $\phi_o \cong \phi_1(1+\varepsilon)$ which, introduced in eq.(3.1.b), gives

$$\phi_o \, (1 - \gamma/2) \cong \phi_o / (1+\varepsilon)$$

from which one gets $\varepsilon \cong \gamma / (2 - \gamma)$. By consequence the first beam results to be

$$\psi_1 \cong \phi_o \, (1 - 2\varepsilon) = \phi_o (2 - 3\gamma) / (2 - \gamma).$$

Substituting in eq.(3.1.c) we obtain

$$f_N(r) \cong \sigma \, \gamma(r) E_o \, \phi_o \, [1 - (2 - 3\gamma)(2-\gamma)^{-1} \sum_{n=0}^{\infty} \gamma^n] / c$$

which, substituting the sum $1/(1-\gamma)$ of the series, gives the force

$$f_N(r) \cong \sigma \gamma^3(r) E_o \, \phi_o / c \, .$$

Substituting $p_M$ and the shielding angle, we obtain the force between two nucleons due to the cosmic radiation pressure

(3.1.d) $$f_N(r) \cong 4\sigma p_M \, (\sigma / 2\pi \, r^2)^3 .$$

It is noteworthy that the nuclear force depends only on the *mutual shielding* of nucleons immersed in the flux of cosmic quanta.

The above formulation must satisfy the preceding rules about nuclear forces when the cosmic pressure $p_M$ takes the right figure. For instance, the nuclear force $f_N(r)$ of a proton upon the nearest proton whose distance $r$ equals the De Broglie's wavelength of relativistic nucleons $\lambda_p \approx 2 \times 10^{-16}$ (see eq.2.2.h, where $v \approx 0.707c$) must be well higher than the electric Coulomb's force. Conversely the nuclear force between two protons at distance $2\lambda_p$ is of the same order of the electric repulsion force (indifferent link). Applying this last condition one gets

(3.1.e) $$4\sigma (\sigma / 8\pi \, \lambda_p^2)^3 p_M \approx e^2 / 16\pi\varepsilon_o \, \lambda_p^2$$

which requires a cosmic pressure $p_M \approx 10^{61}$.
The successive rule affirms that $f_N(r)$ must vanish outside the nucleus. For instance, the force $f_N(r)$ of a nucleus upon the nearest electron results negligible (about $10^{-20}$ times the electric force).

### b) Confinement of nucleons within the atomic nuclei

A more accurate cosmic pressure may be calculated from the stable confinement of nuclei.
Let's consider a heavy nucleus constituted of orbiting nucleons, namely Z protons and N neutrons, confined by the nuclear force given by eq.(3.1.d). The role of the neutrons on the nuclear stability is fundamental because to oppose the Coulomb's repulsion an increasing ratio $N/(Z+N)$ is needed when the number of protons increases. Let's consider the forces on the peripheral proton orbiting at a distance $r_N$ from the centre of nucleus. Each nucleon at a distance $r_k$ exerts a drawing force $f_N(r_k)$ on this proton. Conversely the $(Z-1)$ protons exert a central repulsion force.

Considering the quantised centrifugal force, the peripheral proton satisfies the equilibrium equation

(3.1.f) $$4 \sum_{k=1}^{N+Z-1} \sigma^4 p_M \cos\alpha_k \, /(2\pi \, r_k^2)^3 \cong j(j+1)\hbar^2 / m r_N^3 + (Z-1) \, e^2 / 4\pi\varepsilon_o \, r_N^2.$$

where $r_k = k \lambda_p$ ($k = 1, 2, 3,..$). The peripheral proton is tightly linked to the nearest nucleons.

The strongest nuclear decay is not due to the expulsion of single protons, but of the $\alpha$ particles constituted by 2 protons and 2 neutrons, which represent a fundamental sub-cluster of nucleons. By this reason we assume that the peripheral proton is *tightly* linked to 3 nucleons, whereas the remaining nucleons contribute a little to the proton confinement. Under these assumptions, the above equation reduces to

$$(3.1.g) \qquad 4 \left( \sum_{k=1}^{3} \cos\alpha_k + 0.38 \right) \sigma^4 p_M / 8\pi^3 \lambda_p^6 \approx j(j+1)\hbar^2/mr_N^3 + (Z-1) e^2/4\pi\varepsilon_0 r_N^2 .$$

Solving this equation needs assumptions about the spatial distribution of the 3 nucleons.

Tentatively, the cosmic radiation pressure which makes stable the heavy nuclei ($r_N \approx 1.2 \times 10^{-15}$, $j = 4$) results to be $p_M \approx 10^{61} [1 + (Z-1)/483]$. The substitution $Z = 92$ gives the cosmic pressure which assures the stability of the uranium nucleus (making unstable the transuranic nuclei) $p_M \approx 1.2 \times 10^{61}$. This figure agrees with the indication of eq.(3.1.e).

### 3.2 – The characteristic constants of cosmic quanta and the Planck's length

The constants of the flux of cosmic quanta may be estimated substituting $p_M = E_o \phi_o / 4 c$ into eq.(2.3.e), thus obtaining

$$(3.2.a) \qquad K_o = \pi G / 2 p_M A_o^2$$

from which it is possible to calculate the constants $E_o = K_o mc^2$, $\ni_o = 4 p_M$, $\phi_o = 4 c p_M/E_o$, $h_o = c^2/A_o\phi_o$, $\nu_o = E_o/h_o$, $\lambda_o = c/\nu_o$ when the figure of $A_o$ is known. In Sec.2.4 an approximate value has been calculated ($A_o \approx 4.7 \times 10^{-11}$) through an application to the Earth mass.

The dimensional analysis looking for a physical quantity with dimensions $[A_o] = [m^2/\text{kg}]$ and depending on $h$, $G$, $c$ and on the void density $\ni_o$, gives the following expression

$$(3.2.b) \qquad A_o = (c^3/hG)^{1/2} (c^2/\ni_o) = c^2/4 p_M l_P$$

where $l_P = (hG/c^3)^{1/2} = 4.0518 \times 10^{-35}$ is the Planck's length.

Substituting the figure $p_M \approx 1.2 \times 10^{61}$ one obtains $A_o \approx 4.63 \times 10^{-11}$ which agrees with the number calculated in Sec.2.4. These two figures enable us to evaluate the constants

$$K_o \approx 3.94 \times 10^{-51} \qquad E_o \approx 5.91 \times 10^{-61} \qquad \ni_o \approx 4.8 \times 10^{61}$$
$$\phi_o \approx 2.43 \times 10^{130} \qquad h_o \approx 7.861 \times 10^{-101} \qquad \nu_o \approx 7.50 \times 10^{42} .$$

In particular the cosmic quanta wavelength $\lambda_o = c/\nu_o = c^2/4 p_M A_o$ is given by the same eq.(3.2.b) which defines the Planck's length $l_P$. It is noteworthy that the equality $l_P = \lambda_o$ comes also from the equivalence mass-energy (eq.2.2.e) written as follows $A_o = \nu_o c^2/E_o\phi_o = c^3/E_o\phi_o\lambda_o = c^2/4 p_M\lambda_o$.

The mutual proof between the calculated constants and those derived from dimensional analysis appears to give reliability to the above figures.

Let's finally mention an hypothesis about the structure of the elementary particles linked to the paradigm of the cosmic quanta. If the usual mass-energy equivalence holds for cosmic quanta, the mass density of the "void" ($\delta_o = \ni_o / c^2 \approx 5.3 \times 10^{44}$) is much higher than the apparent nucleon density $\delta_n \approx m/\sigma^{3/2} \approx 8 \times 10^{28}$. This fact would support the hypothesis that the elementary particles may be hollow shells or vortices of a perfect fluid constituted of sub-particles considered massless and collisionless in a paper of K.Tod[28].

An other paper by J.Barrow et al.[29] finds that the stability of the static universe is not significantly changed by the presence of a self-interacting particles of a scalar field source.

**Final remarks**

The present paradigm of the cosmic flux, structured by quanta with proper energy, wavelength,

flux and collision cross-section, appears to hold the conceptual basis for developing a coherent model of the cosmology based on a static-evolving universe, which shows the Hubble's redshift as a consequence of the universe evolution. A second work is in progress.